\documentclass[12pt]{iopart}
\usepackage{iopams,graphicx,citesort}

\def\ii{{\mathrm{i}}}

\def\ee{{\mathrm{e}}}

\def\HH{\mathrm{H}}
\def\VV{\mathrm{V}}

\def\no{{\nonumber}} %\nonumber

\def\bra#1{\langle #1|}

\def\ket#1{|#1\rangle}

\def\bracket#1{\langle #1 \rangle}
\def\bracketi#1#2{\langle #1 | #2 \rangle}
\def\bracketii#1#2#3{\langle #1 | #2| #3\rangle}

\def\sub#1{_\mathrm{#1}} %subscript
\def\tr{\mathrm{tr}}

\begin{document}

\title{A framework for measuring weak values without weak interactions and its diagrammatic representation}

% Of weak-value measurement methods using qubit probes}

\author{Kazuhisa Ogawa$^1$, Osamu Yasuhiko$^2$, Hirokazu Kobayashi$^3$, Toshihiro Nakanishi$^2$, and Akihisa Tomita$^1$}
\address{$^1$Graduate School of Information Science and Technology, Hokkaido University, Kita 14, Nishi 9, Kita-ku, Sapporo 060-0814, Japan\\
$^2$Department of Electronic Science and Engineering, Kyoto University, Kyoto daigaku-katsura, Nishikyo-ku, Kyoto 615-8510, Japan\\
$^3$School of System Engineering, Kochi University of Technology, Miyanoguchi 185, Tosayamada, Kami City, Kochi 782-8502, Japan}
\ead{ogawak@ist.hokudai.ac.jp}
\vspace{10pt}
% \begin{indented}
% \item[]July 2018
% \end{indented}

\begin{abstract}
Weak values are typically obtained experimentally by performing weak measurements, which involve weak interactions between the measured system and a probe.
However, the determination of weak values does not necessarily require weak measurements, and several methods without weak system--probe interactions have been developed previously.
In this work, a framework for measuring weak values is proposed to describe the relationship between various weak measurement techniques in a unified manner.
This framework, which uses a probe-controlled system transformation instead of the weak system--probe interaction, improves the understanding of the currently used weak value measurement methods.
Furthermore, a diagrammatic representation of the proposed framework is introduced to intuitively identify the complex values obtained in each measurement system. 
By using this diagram, a new method for measuring weak values with a desired function can be systematically derived. 
As an example, a scan-free and more efficient direct measurement method of wavefunctions than the conventional techniques using weak measurements is developed. 
\end{abstract}

%Uncomment for keywords
\vspace{2pc}

%Uncomment for Submitted to journal title message
%\submitto{\NJP}

%Uncomment if a separate title page is required
\maketitle

%For two-column output uncomment the next line and choose [10pt] rather than [12pt] in the \documentclass declaration
%\ioptwocol

\section{Introduction}

The concept of weak values was first introduced by Aharonov \textit{et al.} \cite{PhysRevLett.60.1351} as observable statistics influenced not only by the initial state $\ket{\psi\sub{i}}$, but also by the final state $\ket{\psi\sub{f}}$ of the studied system.
The weak value of the observable $\hat{A}$ is defined as $\bracket{\hat{A}}\sub{w}:=\bracketii{\psi\sub{f}}{\hat{A}}{\psi\sub{i}}/\bracketi{\psi\sub{f}}{\psi\sub{i}}$, which is generally a complex number.
Weak values are typically determined experimentally using a technique called weak measurements.
In weak measurements, a probe interacts with the system weakly enough not to significantly disturb the system. 
After the post-selection of the system's final state, a weak value is obtained as the average of the measurement outcomes of the probe. 
Weak values have been used to study various fundamental problems in quantum mechanics \cite{PhysRevLett.102.020404,yokota2009direct,lund2010measuring,kocsis2011observing,goggin2011violation,rozema2012violation,denkmayr2014observation,kaneda2014experimental,mahler2016experimental} because they represent intermediate aspects of pre- and post-selected systems. 
In addition, weak values are used for the precise measurement of the magnitudes of weak system--probe interactions \cite{hosten2008observation,dixon2009ultrasensitive,magana2014amplification,hallaji2017weak} because they can exceed the eigenvalues of the observables.
Furthermore, weak values are utilized for the direct measurement of complex functions such as wavefunctions and pseudo-probability distributions of the initial state of the system \cite{lundeen2011direct,lundeen2012procedure,salvail2013full,kobayashi2014stereographical,malik2014direct,mirhosseini2014compressive,shi2015scan,thekkadath2016direct,piacentini2016measuring,bolduc2016direct}.% by conforming them to such complex functions. 

In previous studies, weak values have been obtained by weak measurements, which involve weak system--probe interactions.
However, this is not the only possible method for their determination, and several alternative measurement techniques that do not involve weak system--probe interactions have been recently developed, including those using strong system--probe interactions \cite{zou2015direct,vallone2016strong,denkmayr2017experimental,calderaro2018direct}, modular values \cite{kedem2010modular}, quantum control interactions \cite{hofmann2014sequential}, an enlarged Hilbert space \cite{ho2018quantum}, and coupling-deformed pointer observables \cite{zhang2016coupling}.
In this work, we propose a framework for measuring weak values common to the conventional weak measurement using a qubit probe and other weak value measurement techniques.
This framework uses a probe-controlled system transformation instead of the weak system--probe interaction to obtain a unified description of the relationship between these measurement methods.
In addition, a diagrammatic representation of the proposed framework is introduced.
It illustrates the time evolution of the system's state for each probe mode and allows us to intuitively identify the complex values obtained in each measurement system.
Furthermore, using the transformation rule of the diagram, a new method for measuring weak values that has a desired function can be systematically derived.
As an example, this transformation rule is applied to a direct measurement method of wavefunctions using weak measurement \cite{lundeen2011direct} to develop a scan-free and more efficient method than the conventional techniques using weak measurements \cite{lundeen2011direct,lundeen2012procedure,salvail2013full,kobayashi2014stereographical,malik2014direct,mirhosseini2014compressive,shi2015scan,thekkadath2016direct,piacentini2016measuring,bolduc2016direct}.

\section{A framework for measuring weak values without weak interactions---definitions and general properties}\label{sec:meas-fram-weak}

In this section, we first present a framework for measuring weak values without weak system--probe interactions and construct a diagram that intuitively describes the relationship between the time evolution of the system and complex values obtained by the measurements.
Next, we formulate a transformation rule of the diagram, which can be used to derive another weak value measurement method that produces the same results.

For this purpose, we consider the composite system $\mathcal{H}\sub{S}\otimes\mathcal{H}\sub{P}$ of the measured system $\mathcal{H}\sub{S}$ (arbitrary dimension) and the qubit probe $\mathcal{H}\sub{P}$.
Pauli operators in $\mathcal{H}\sub{P}$ with the orthonormal basis $\{\ket{0},\ket{1}\}$ can be expressed as
\begin{eqnarray}
\hat{\sigma}_x&=\ket{0}\bra{1}+\ket{1}\bra{0}=\ket{+}\bra{+}-\ket{-}\bra{-},\\
\hat{\sigma}_y&=-\ii\ket{0}\bra{1}+\ii\ket{1}\bra{0}=\ket{\!+\ii}\bra{+\ii}-\ket{\!-\ii}\bra{-\ii},\\
\hat{\sigma}_z&=\ket{0}\bra{0}-\ket{1}\bra{1},
\end{eqnarray}
where $\ket{\pm}:=(\ket{0}\pm\ket{1})/\sqrt{2}$, $\ket{\!\pm\ii}:=(\ket{0}\pm\ii\ket{1})/\sqrt{2}$.

\subsection{ Measurement framework and its diagrammatic representation}\label{sec:theor-fram-its}

\begin{figure}
\begin{center}
\includegraphics[width=12cm]{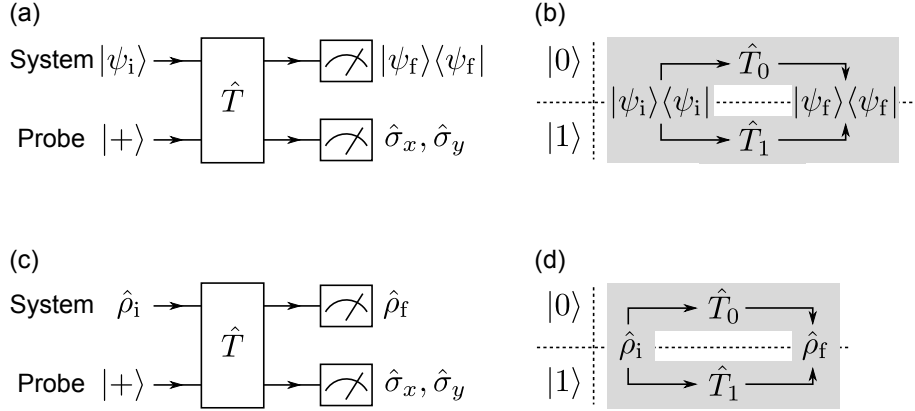}
\caption{
(a) Quantum circuit of the proposed measurement framework of weak values.
(b) Diagram describing circuit (a) as the time evolution of the system for each probe mode.
(c) Quantum circuit of panel (a), in which the pre- and post-selected states correspond to the general (mixed) states $\hat{\rho}\sub{i}$ and $\hat{\rho}\sub{f}$, respectively.
(d) Diagram representing circuit (c).
 }\label{fig:1}
\end{center}
\end{figure}

% Weak measurement using a qubit probe and many of the weak-value measurement methods \cite{zou2015direct,vallone2016strong,denkmayr2017experimental,calderaro2018direct,kedem2010modular,hofmann2014sequential,ho2018quantum} are interpreted as
Let us consider the measurement framework shown in Fig.~\ref{fig:1}(a).
In this system, the initial state is set as $\ket{\!\mathit{\Psi}\sub{i}}=\ket{\psi\sub{i}}\ket{+}$.
It first undergoes the following probe-controlled system transformation $\hat{T}$:
\begin{eqnarray}
\hat{T}:=\hat{T}_0\otimes\ket{0}\bra{0}+\hat{T}_1\otimes\ket{1}\bra{1},\label{eq:2}
\end{eqnarray}
where the absolute values of the eigenvalues of $\hat{T}_0$ and $\hat{T}_1$ are assumed not to exceed one, so that these transformations are physically realizable. 
If $\hat{T}_0$ and $\hat{T}_1$ are unitary operators, $\hat{T}$ is also a unitary operator; otherwise, the initial state undergoes the transformation $\hat{T}$ with a probability of less than one.
The unnormalized state after the transformation $\hat{T}$ is described as $\hat{T}\ket{\!\mathit{\Psi}\sub{i}}$, which means that the normalized state $\hat{T}\ket{\!\mathit{\Psi}\sub{i}}/\|\hat{T}\ket{\!\mathit{\Psi}\sub{i}}\|$ is obtained with the probability $\|\hat{T}\ket{\!\mathit{\Psi}\sub{i}}\|^2$.
Next, the system is projected onto $\ket{\psi\sub{f}}$, and the probe is measured in the bases $\{\ket{\pm}\}$ and $\{\ket{\!\pm\ii}\}$.
Let $P(\pm)$ [$P(\pm\ii)$] be the probabilities that the initial state $\ket{\!\mathit{\Psi}\sub{i}}$ undergoes the transformation $\hat{T}$, the system is projected onto $\ket{\psi\sub{f}}$, and the probe is projected onto $\ket{\pm}$ [$\ket{\!\pm\ii}$].
$P(\pm)$ and $P(\pm\ii)$ are expressed as
\begin{eqnarray}
P(\pm)&=
\bra{\!\mathit{\Psi}\sub{i}}\hat{T}^\dag
\left(\ket{\psi\sub{f}}\bra{\psi\sub{f}}\otimes\ket{\pm}\bra{\pm}\right)
\hat{T}\ket{\!\mathit{\Psi}\sub{i}}
\no\\
&=\frac{1}{4}
\left(
|\bracketii{\psi\sub{f}}{\hat{T}_0}{\psi\sub{i}}|^2
+|\bracketii{\psi\sub{f}}{\hat{T}_1}{\psi\sub{i}}|^2
\pm 2\mathrm{Re}\bracketii{\psi\sub{i}}{\hat{T}_0^\dag\ket{\psi\sub{f}}\bra{\psi\sub{f}}\hat{T}_1}{\psi\sub{i}}
\right),\\
P(\pm\ii)&=
\bra{\!\mathit{\Psi}\sub{i}}\hat{T}^\dag
\left(\ket{\psi\sub{f}}\bra{\psi\sub{f}}\otimes\ket{\!\pm\ii}\bra{\pm\ii}\right)
\hat{T}\ket{\!\mathit{\Psi}\sub{i}}
\no\\
&=\frac{1}{4}
\left(
|\bracketii{\psi\sub{f}}{\hat{T}_0}{\psi\sub{i}}|^2
+|\bracketii{\psi\sub{f}}{\hat{T}_1}{\psi\sub{i}}|^2
\pm 2\mathrm{Im}\bracketii{\psi\sub{i}}{\hat{T}_0^\dag\ket{\psi\sub{f}}\bra{\psi\sub{f}}\hat{T}_1}{\psi\sub{i}}
\right).
\end{eqnarray}
By measuring $P(\pm)$ and $P(\pm\ii)$ experimentally, the complex value $\bracketii{\psi\sub{i}}{\hat{T}_0^\dag\ket{\psi\sub{f}}\bra{\psi\sub{f}}\hat{T}_1}{\psi\sub{i}}$ is directly derived as
\begin{eqnarray}
P(+)-P(-)+\ii[P(+\ii)-P(-\ii)]
=\bracketii{\psi\sub{i}}{\hat{T}_0^\dag\ket{\psi\sub{f}}\bra{\psi\sub{f}}\hat{T}_1}{\psi\sub{i}}.\label{eq:7}
\end{eqnarray}
If $\{\hat{T}_0, \hat{T}_1\}$ is selected to be $\{\hat{1},\hat{A}\}$, the derived complex value becomes $\bracketi{\psi\sub{i}}{\psi\sub{f}}\bracketii{\psi\sub{f}}{\hat{A}}{\psi\sub{i}}$, which is proportional to the weak value $\bracket{\hat{A}}\sub{w}$.
Thus, the proposed framework can be used for measuring weak values.
The detailed relationship between this framework and other weak value measurement techniques (including those without weak interactions) will be explained in Sec.~\ref{sec:underst-weak-value}.

The time evolution of the quantum system described in Fig.~\ref{fig:1}(a) can be illustrated by the diagram shown in Fig.~\ref{fig:1}(b).
In this diagram, the upper and lower rows represent the time evolutions of the system in the probe modes $\ket{0}$ and $\ket{1}$, respectively.
The initial and final states of the system are described by the density operators $\ket{\psi\sub{i}}\bra{\psi\sub{i}}$ and $\ket{\psi\sub{f}}\bra{\psi\sub{f}}$, respectively.
For the probe, it is assumed that the initial state $\ket{+}$ is set at the left end, and the expectation values of $\hat{\sigma}_x$ and $\hat{\sigma}_y$ are measured at the right end.  
The complex value derived using Eq.~(\ref{eq:7}) is $\bracketii{\psi\sub{i}}{\hat{T}_0^\dag\ket{\psi\sub{f}}\bra{\psi\sub{f}}\hat{T}_1}{\psi\sub{i}}=\tr(\ket{\psi\sub{i}}\bra{\psi\sub{i}}\hat{T}_0^\dag\ket{\psi\sub{f}}\bra{\psi\sub{f}}\hat{T}_1)$, which is the trace of the product of the operators $\ket{\psi\sub{i}}\bra{\psi\sub{i}}$, $\hat{T}_0^\dag$, $\ket{\psi\sub{f}}\bra{\psi\sub{f}}$, and $\hat{T}_1$.
These operators are ordered clockwise along the loop in Fig.~\ref{fig:1}(b); therefore, the derived complex value can be intuitively identified as the trace of the clockwise product of the operators along the loop.
Note that $\hat{T}_0$ must be an Hermitian conjugate operator.

This diagram can be modified for the system in which pre- and post-selected states are the mixed states $\hat{\rho}\sub{i}$ and ${\hat{\rho}\sub{f}}$, respectively, as shown in Fig.~\ref{fig:1}(c).
In this case, the derived complex value is expressed as $\tr(\hat{\rho}\sub{i}\hat{T}_0^\dag\hat{\rho}\sub{f}\hat{T}_1)$, which is also identified in the diagram of Fig.~\ref{fig:1}(d) as the trace of the clockwise product of the operators $\hat{\rho}\sub{i}$, $\hat{T}_0^\dag$, $\hat{\rho}\sub{f}$, and $\hat{T}_1$ along the loop.
% as the trace of the product created by connecting the operators on the loop in the diagram 

\subsection{Transformation rules of the diagram}\label{sec:transf-rules-diagr}

\begin{figure}
\begin{center}
\includegraphics[width=15cm]{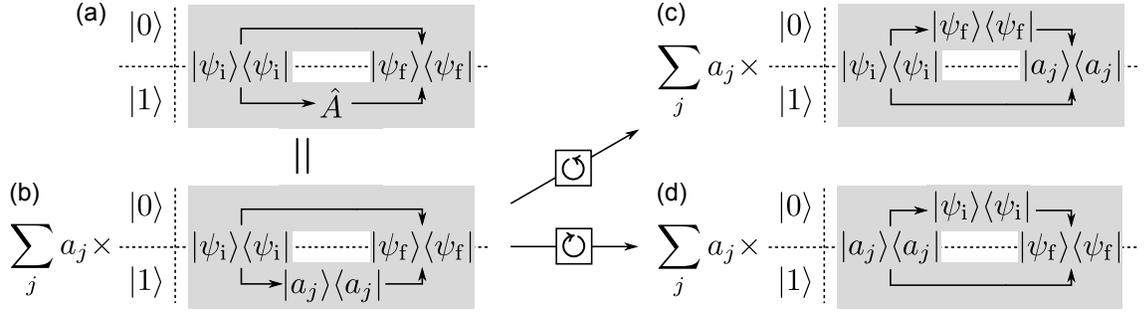}
\caption{
(a) Diagram describing the time evolution of the weak value measurement system.
(b) Measurement system obtained by spectrally decomposing $\hat{A}$ in diagram (a).
(c) Measurement system obtained by rotating the loop in diagram (b) counterclockwise.
% If $\hat{A}$ is a non-negative operator, it can be considered to perform generalized measurement with a POVM element $k\hat{A}$, where $k$ is an appropriate coefficient $k$.
(d) Measurement system obtained by rotating the loop in diagram (b) clockwise.
% If $\hat{A}$ is a non-negative operator, it can be considered to prepare an initial state represented by the density operator $k\hat{A}$.
}\label{fig:3}
\end{center}
\end{figure}

In the measurement framework described above, the derived complex value can be identified as the trace of the clockwise product of the operators along the loop in the diagram.
Due to the cyclic property of the trace $\tr(\hat{A}\hat{B})=\tr(\hat{B}\hat{A})$, this value is invariant under cyclic permutations of the operators along the loop. 
We call this property the diagram transformation rule.
In the following, we show an example of transformation of the weak value measurement system using this transformation rule.

Let us consider the time evolution described by the diagram depicted in Fig.~\ref{fig:3}(a).
It shows that the derived complex value is $\tr(\ket{\psi\sub{i}}\bracketi{\psi\sub{i}}{\psi\sub{f}}\bra{\psi\sub{f}}\hat{A})=|\bracketi{\psi\sub{f}}{\psi\sub{i}}|^2\bracket{\hat{A}}\sub{w}$.
Other measurement systems producing the same complex value can be systematically derived using the transformation rule of the diagram as follows.

First, if $\hat{A}$ is spectrally decomposed as $\hat{A}=\sum_ja_j\ket{a_j}\bra{a_j}$, the obtained complex value can be expressed as $\bracketi{\psi\sub{i}}{\psi\sub{f}}\bracketii{\psi\sub{f}}{\hat{A}}{\psi\sub{i}}=\sum_ja_j\bracketi{\psi\sub{i}}{\psi\sub{f}}\bracketi{\psi\sub{f}}{a_j}\bracketi{a_j}{\psi\sub{i}}$.
Therefore, the measurement system yielding the same complex value can be described by Fig.~\ref{fig:3}(b), and the complex value $\bracketi{\psi\sub{i}}{\psi\sub{f}}\bracketii{\psi\sub{f}}{\hat{A}}{\psi\sub{i}}$ can be constructed by obtaining $\bracketi{\psi\sub{i}}{\psi\sub{f}}\bracketi{\psi\sub{f}}{a_j}\bracketi{a_j}{\psi\sub{i}}$ for each $j$ and adding these values multiplied by weights $a_j$.

Next, after cyclically moving the operators in Fig.~\ref{fig:3}(b) along the loop counterclockwise [clockwise], the diagrams depicted in Fig.~\ref{fig:3}(c) [(d)] can be obtained.
In the diagram of Fig.~\ref{fig:3}(c), the system originally set as $\ket{\psi\sub{i}}$ is first projected onto $\ket{\psi\sub{f}}$ in the probe mode $\ket{0}$, and then onto $\ket{a_j}$ in the both probe mode. 
The derived complex value is $\bracketi{\psi\sub{i}}{\psi\sub{f}}\bracketi{\psi\sub{f}}{a_j}\bracketi{a_j}{\psi\sub{i}}$, which corresponds to the time evolution of Fig.~\ref{fig:3}(b).
If $\hat{A}$ is a non-negative operator, the final measurement can be regarded as generalized measurement represented by the positive-operator valued measure (POVM) element $k\hat{A}$, where $k$ is an appropriate coefficient.
On the other hand, in the diagram depicted in Fig.~\ref{fig:3}(d), the system originally set as $\ket{a_j}$ is first projected onto $\ket{\psi\sub{i}}$ in the probe mode $\ket{0}$, and then onto $\ket{\psi\sub{f}}$ in the both probe mode; the derived complex value corresponds to the time evolution described in Fig.~\ref{fig:3}(b) and (c).
If $\hat{A}$ is a non-negative operator, the state preparation can be replaced with the preparation of the state represented by the density operator $k\hat{A}$.

\section{Understanding weak value measurement methods in the proposed framework}\label{sec:underst-weak-value}

The proposed measurement framework can be used to better describe the conventional weak measurement and other weak value measurement methods in a unified manner.
In this section, we explain how these measurement methods can be understood in the proposed framework.

\subsection{Weak measurements in the framework}\label{sec:3}

Weak measurements using a qubit probe can be included in the measurement framework described above with a small modification.
Here, we first review the conventional weak measurement using a qubit probe and then explain how they can be modified to fit the framework.

\begin{figure}
\begin{center}
\includegraphics[width=17cm]{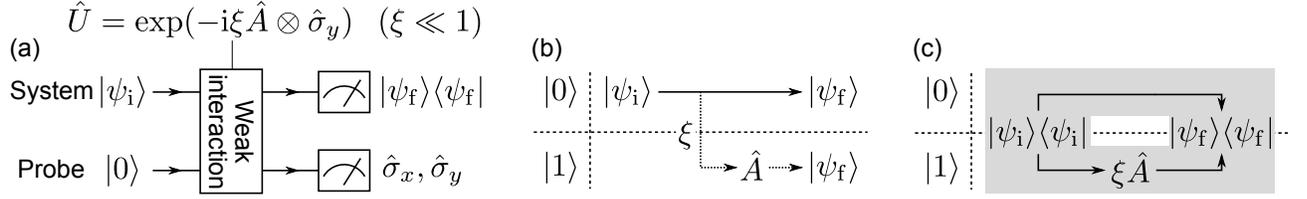}
\caption{
(a) Quantum circuit used for weak measurements with a qubit probe.
(b) Diagram describing the time evolution of the system for each probe mode in panel (a).
(c) Diagram describing the time evolution of the modified weak measurement system.
 }\label{fig:2}
\end{center}
\end{figure}

The conventional weak measurement procedure using a qubit probe can be described by the quantum circuit depicted in Fig.~\ref{fig:2}(a).
The initial states of the system and the probe are $\ket{\psi\sub{i}}$ and $\ket{0}$, respectively.
First, the initial state $\ket{\psi\sub{i}}\ket{0}$ is evolved through the weak system--probe interaction $\exp(-\ii\xi\hat{A}\otimes\hat{\sigma}_y)$, where $\hat{A}\otimes\hat{\sigma}_y$ is an interaction Hamiltonian, and $\xi$ is a small coupling constant $(\xi\ll 1)$. It is further transformed into
\begin{eqnarray}
\exp(-\ii\xi\hat{A}\otimes\hat{\sigma}_y)\ket{\psi\sub{i}}\ket{0}
&=\left[\hat{1}-\ii\xi\hat{A}\otimes\hat{\sigma}_y+O(\xi^2)\right]\ket{\psi\sub{i}}\ket{0}\no\\
&=\ket{\psi\sub{i}}\ket{0}+\xi\hat{A}\ket{\psi\sub{i}}\ket{1}+O(\xi^2).\label{eq:1}
\end{eqnarray}
Since this transformation is unitary, the state after the interaction is obtained with a probability of one.
Next, the system is projected onto $\ket{\psi\sub{f}}$, and the probe is measured in the bases $\ket{\pm}$ and $\ket{\!\pm\ii}$.
% Next, the system is post-selected into the final state $\ket{\psi\sub{f}}$.
% The (unstandardized) state $\ket{\psi'}$ of the probe system after the post-selection can be expressed as 
% Finally, the probe state $\ket{\psi'}$ is measured by the projective measurements $\{\ket{\pm}\bra{\pm}\}$ or $\{\ket{\!\pm\ii}\bra{\pm\ii}\}$.
Let $P(\pm)$ [$P(\pm\ii)$] be the probabilities that the system is projected onto $\ket{\psi\sub{f}}$ and the probe is projected onto $\ket{\pm}$ [$\ket{\!\pm\ii}$].
In this case, $P(\pm)$ and $P(\pm\ii)$ can be expressed as
\begin{eqnarray}
P(\pm)&=
|\bra{\psi\sub{f}}\bra{\pm}\exp(-\ii\xi\hat{A}\otimes\hat{\sigma}_y)\ket{\psi\sub{i}}\ket{0}|^2\no\\
&=|\bracketi{\psi\sub{f}}{\psi\sub{i}}|^2
\left(\frac{1}{2}\pm\xi\mathrm{Re}\bracket{\hat{A}}\sub{w}\right)+O(\xi^2),\\
P(\pm\ii)&=
|\bra{\psi\sub{f}}\bra{\pm\ii}\exp(-\ii\xi\hat{A}\otimes\hat{\sigma}_y)\ket{\psi\sub{i}}\ket{0}|^2\no\\
&=|\bracketi{\psi\sub{f}}{\psi\sub{i}}|^2
\left(\frac{1}{2}\pm\xi\mathrm{Im}\bracket{\hat{A}}\sub{w}\right)+O(\xi^2).
\end{eqnarray}
Then, the weak value $\bracket{\hat{A}}\sub{w}$ is described by using $P(\pm)$ and $P(\pm\ii)$ as 
\begin{eqnarray}
P(+)-P(-)+\ii[P(+\ii)-P(-\ii)]=2\xi
|\bracketi{\psi\sub{f}}{\psi\sub{i}}|^2
\bracket{\hat{A}}\sub{w}
+O(\xi^2).\label{eq:8}
\end{eqnarray}
Since $|\bracketi{\psi\sub{f}}{\psi\sub{i}}|^2=P(+)+P(-)+O(\xi^2)$, if $\xi$ is known, the weak value $\bracket{\hat{A}}\sub{w}$ is derived experimentally by measuring $P(\pm)$ and $P(\pm\ii)$.
The time evolution during this weak measurement can be represented by the diagram depicted in Fig.~\ref{fig:2}(b).
According to this diagram, by ignoring $O(\xi^2)$, the weak interaction can be understood as an operation that extracts a small portion of $\ket{\psi\sub{i}}$ proportional to $\xi$ from the probe mode $\ket{0}$, transfers this portion to the mode $\ket{1}$, and multiplies it by $\hat{A}$.

Next, we modify the conventional weak measurement method to fit the proposed measurement framework. 
Instead of using the weak interaction, the state described by Eq.~(\ref{eq:1}) can be produced without $O(\xi^2)$ by setting the initial state $\ket{\!\mathit{\Psi}\sub{i}}=\ket{\psi\sub{i}}\ket{+}$ and transforming it via the probe-controlled system transformation $\hat{T}=\hat{1}\otimes\ket{0}\bra{0}+\xi\hat{A}\otimes\ket{1}\bra{1}$.
The absolute values $\xi a_j$ of the eigenvalues of $\xi\hat{A}=\sum_j\xi a_j\ket{a_j}\bra{a_j}$ do not exceed one so that the transformation $\xi\hat{A}$ is physically realizable.
By performing this modification, the weak measurement system satisfies the developed measurement framework where $\{\hat{T}_0,\hat{T}_1\}=\{\hat{1},\xi\hat{A}\}$ in Eq.~(\ref{eq:2}). 
This time evolution is described by the diagram in Fig.~\ref{fig:2}(c).
Let $P(\pm)$, $P(\pm\ii)$, and $P(0)$ be the probabilities that the initial state $\ket{\!\mathit{\Psi}\sub{i}}$ undergoes the transformation $\hat{T}$, the system is projected onto $\ket{\psi\sub{f}}$, and the probe is projected onto $\ket{\pm}$, $\ket{\pm\ii}$, and $\ket{0}$, respectively.
Using Eq.~(\ref{eq:7}), the weak value $\bracket{\hat{A}}\sub{w}$ can be described via $P(\pm)$, $P(\pm\ii)$, and $P(0)$ as 
\begin{eqnarray}
P(+)-P(-)+\ii[P(+\ii)-P(-\ii)]
&=
\bracketi{\psi\sub{i}}{\psi\sub{f}}\bracketii{\psi\sub{f}}{\xi\hat{A}}{\psi\sub{i}}\no\\
&=2P(0)\xi\bracket{\hat{A}}\sub{w}.\label{eq:6}
\end{eqnarray}
Although the modified weak measurement method requires an additional measurement of $P(0)=|\bracketi{\psi\sub{f}}{\psi\sub{i}}|^2/2$ (unlike the conventional weak measurement technique), if $\xi$ is known, the weak value $\bracket{\hat{A}}\sub{w}$ can be derived experimentally.
The obtained complex value $\bracketi{\psi\sub{i}}{\psi\sub{f}}\bracketii{\psi\sub{f}}{\xi\hat{A}}{\psi\sub{i}}=\tr\left(\ket{\psi\sub{i}}\bracketi{\psi\sub{i}}{\psi\sub{f}}\bra{\psi\sub{f}}\xi\hat{A}\right)$ can be identified using the diagram depicted in Fig.~\ref{fig:2}(c), as was described previously in Sec.~\ref{sec:theor-fram-its}.
 
It should be noted that the modified weak measurement method allows estimation of weak values at higher accuracy and precision as compared to those of the conventional one.
According to Eq.~(\ref{eq:8}), the estimator of weak values in the conventional weak measurement includes the error term $O(\xi)$, which does not disappear even in the limit of a large number of measurements.
On the other hand, in the modified weak measurement method, the estimator of weak values contains no error terms as shown in Eq.~(\ref{eq:6}), and the obtained results match the true weak values in the asymptotic limit; as a result, the modified weak measurement technique exhibits higher accuracy than that of the conventional method.
Moreover, during the conventional weak measurements, $\xi$ must be small enough to reduce the effect of the error term $O(\xi)$, and selecting small $\xi$ values results in a large estimation uncertainty proportional to $\xi^{-1}$.
In the modified weak measurement method, $\xi$ can be as high as $a\sub{max}^{-1}$ ($a\sub{max}$ is the maximum eigenvalue of $\hat{A}$); therefore, the estimation uncertainty can be smaller than that of the conventional method at the same number of measurements.

\subsection{Other weak value measurement methods in the proposed framework}\label{sec:other-meas-meth}

Many other measurement methods of weak values can also fit the proposed measurement framework like weak measurement.
In the following, we mention some examples of them and explain how they correspond to the proposed measurement framework.

When the measured observable $\hat{A}$ corresponds to the projection operator $\hat{A}^2=\hat{A}$ or Pauli operator $\hat{\sigma}_i$ ($i=x,y,z$), the strength of the system--probe interaction can be arbitrarily high \cite{zou2015direct,calderaro2018direct,vallone2016strong,denkmayr2017experimental}.
In the former case \cite{zou2015direct,calderaro2018direct,vallone2016strong}, the interaction with the optimal coupling strength $\xi=\pi/2$, which maximally entangles the total state, can be described as $\exp[-\ii(\pi/2)\hat{A}\otimes\hat{\sigma}_y]=\hat{1}-\hat{A}\otimes(\hat{1}+\ii\hat{\sigma}_y)$, and the resulting state after the interaction is $(\hat{1}-\hat{A})\ket{\psi\sub{i}}\ket{0}+\hat{A}\ket{\psi\sub{i}}\ket{1}$.
Therefore, this measurement method corresponds to the case $\{\hat{T}_0,\hat{T}_1\}=\{\hat{1}-\hat{A},\hat{A}\}$ in our framework.
The resulting complex value is $|\bracketi{\psi\sub{f}}{\psi\sub{i}}|^2(\bracket{\hat{A}}\sub{w}-|\bracket{\hat{A}}\sub{w}|^2)$, and the weak value $\bracket{\hat{A}}\sub{w}$ is derived experimentally by measuring $P(\pm)$, $P(\pm\ii)$, and $P(1)=|\bracketi{\psi\sub{f}}{\psi\sub{i}}|^2|\bracket{\hat{A}}\sub{w}|^2/2$, where $P(1)$ is the probability that the initial state undergoes the transformation $\hat{T}$, the system is projected onto $\ket{\psi\sub{f}}$, and the probe is projected onto $\ket{1}$.
In the latter case \cite{denkmayr2017experimental}, the interaction with the optimal coupling strength $\xi=\pi/4$ can be described as $\exp[-\ii(\pi/4)\hat{\sigma}_i\otimes\hat{\sigma}_y]=(\hat{1}-\ii\hat{\sigma}_i\otimes\hat{\sigma}_y)/\sqrt{2}$, and the resulting state after the interaction is $(\ket{\psi}\ket{0}+\hat{\sigma}_i\ket{\psi}\ket{1})/\sqrt{2}$.
This measurement method corresponds to the case $\{\hat{T}_0,\hat{T}_1\}=\{\hat{1},\hat{\sigma}_i\}$ in our framework, and the resulting complex value is $|\bracketi{\psi\sub{f}}{\psi\sub{i}}|^2\bracket{\hat{\sigma}_i}\sub{w}$, which includes the weak value $\bracket{\hat{\sigma}_i}\sub{w}$.

When the measured observable $\hat{A}$ is arbitrary and the strength of the system--probe interaction $\xi$ is arbitrarily high, a modular value is used as the parameter of a pre- and post-selected quantum systems (instead of the weak value) that provides a complete description of its effect on the qubit probe \cite{kedem2010modular}.
When the initial state is $\ket{\psi\sub{i}}\ket{+}$, the interaction is $\exp(-\ii\xi\hat{A}\otimes\ket{1}\bra{1})$, and the post-selected state of the system is $\ket{\psi\sub{f}}$, the resulting unnormalized probe state is $\ket{0}+\bracket{\hat{A}}\sub{m}\ket{1}$, where $\bracket{\hat{A}}\sub{m}:=\bracketii{\psi\sub{f}}{\ee^{-\ii \xi\hat{A}}}{\psi\sub{i}}/\bracketi{\psi\sub{f}}{\psi\sub{i}}$ is a modular value.
The modular value is obtained by measuring the probe state in the bases $\{\ket{\pm}\}$ and $\{\ket{\!\pm\ii}\}$.
This measurement method corresponds to the case $\{\hat{T}_0,\hat{T}_1\}=\{\hat{1},\ee^{-\ii \xi\hat{A}}\}$ in our measurement framework, and the resulting complex value is $|\bracketi{\psi\sub{f}}{\psi\sub{i}}|^2\bracket{\hat{A}}\sub{m}$.
When $\xi=\pi/2$ and $\hat{A}$ is a Pauli operator, $\bracket{\hat{A}}\sub{m}=-\ii\bracket{\hat{A}}\sub{w}$, and the modular value coincides with the weak value of $\hat{A}$. 

In Ref.~\cite{hofmann2014sequential}, the probe-controlled system transformation $\hat{T}$ is used to investigate the trade-off between the measurement back-action and the resolution.
When $\{\hat{T}_0,\hat{T}_1\}=\{\hat{1},\ket{a}\bra{a}\}$, and the initial and final states of the system are $\hat{\rho}\sub{in}$ and $\ket{b}\bra{b}$, respectively, the probe measurement results in various bases reflect the effect of the successive measurements $\ket{a}\bra{a}$ and $\ket{b}\bra{b}$ for $\hat{\rho}\sub{in}$.
In particular, a part of the measurement results exhibits the Kirkwood-Dirac distribution $\rho(a,b):=\bracketi{b}{a}\bracketii{a}{\hat{\rho}\sub{in}}{b}$, which is a complex joint probability distribution showing a nonclassical correlation between the non-commuting operators $\ket{a}\bra{a}$ and $\ket{b}\bra{b}$.
In our measurement framework, $\rho(a,b)$ can be obtained via Eq.~(\ref{eq:7}) directly, while it has been previously determined by weak measurements \cite{PhysRevLett.112.070405} as a numerator of $\bracket{\ket{a}\bra{a}}\sub{w}=\tr(\hat{\rho}\sub{in}\ket{b}\bracketi{b}{a}\bra{a})/\tr(\hat{\rho}\sub{i}\ket{b}\bra{b})$, which is the weak value of $\ket{a}\bra{a}$ for the initial state $\hat{\rho}\sub{in}$ and final state $\ket{b}\bra{b}$.

The weak value measurement method using an expanded Hilbert space \cite{ho2018quantum} is derived by modulating the transformed weak value measurement technique described in Fig.~\ref{fig:3}(c) of Sec.~\ref{sec:transf-rules-diagr}.
The complex value derived by this method is $\bracketi{\psi\sub{i}}{\psi\sub{f}}\bracketii{\psi\sub{f}}{\hat{A}}{\psi\sub{i}}$, but determination of $\bracketi{\psi\sub{i}}{\psi\sub{f}}$ in this value is not required for obtaining the weak value $\bracket{\hat{A}}\sub{w}$.
Therefore, we can omit the projection onto $\ket{\psi\sub{f}}$ in the upper part of the diagram in Fig.~\ref{fig:3}(c); instead, it is necessary only to set the total initial state $(\ket{\psi\sub{f}}\ket{0}+\ket{\psi\sub{i}}\ket{1})/\sqrt{2}$.
This modification produces a weak value measurement method using the expanded Hilbert space.

\section{Application of the diagram---derivation of a new direct measurement method of wavefunctions}

In Sec.~\ref{sec:meas-fram-weak}, we introduced a diagram to intuitively identify complex values obtained in our measurement framework and its transformation rule.
Both the diagram and the transformation rule can be used for the systematical derivation of a new measurement method of weak values that has a desired function.
In this section, the derivation procedure is described in more detail.
As an example, here we deal with the direct measurement of wavefunctions by weak measurement \cite{lundeen2011direct};
we develop a scan-free and more efficient direct measurement method of wavefunctions than the conventional techniques using weak measurements.

\begin{figure}
\begin{center}
\includegraphics[width=17cm]{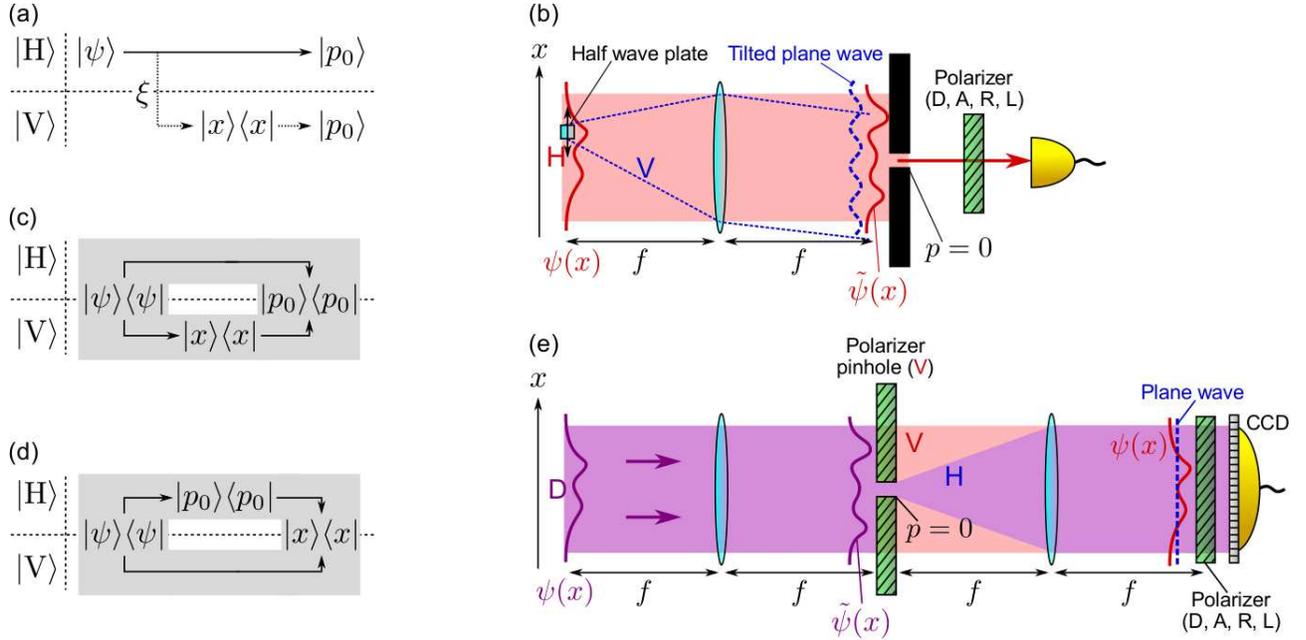}
\caption{
(a) Diagram of the time evolution of the direct measurement of wavefunctions by weak measurements.
(b) An optical system used for the realization of the time evolution process described in panel (a).
On the Fourier plane just before the slit extracting the $p=0$ component, the Fourier transform of the original wavefunction $\tilde{\psi}(x)$ and tilted plane-wave self-reference light appear in each polarization mode.% ; however, it is unclear how these interferences will yield the direct measurement of the wavefunction $\psi(x)$.
(c) Diagram of the time evolution described in panel (a) transformed to fit the proposed measurement framework.
(d) Diagram illustrating the method derived by modifying diagram (c) so that the direct measurement of the wavefunction $\psi(x)$ is performed more efficiently without scanning.
(e) An optical system that realizes the measurement method described in diagram (d).
On the image plane just before the polarizer and detector, the original wavefunction $\psi(x)$ and non-tilted plane-wave self-referencing light appear in each polarization mode.% ; therefore, it is clear that interference at each position $x$ yields the direct measurement of the wavefunction $\psi(x)$.
}\label{fig:4}
\end{center}
\end{figure}

The direct measurement of wavefunctions by weak measurements has been first proposed and experimentally realized by Lundeen \textit{et al.}~\cite{lundeen2011direct}, where ``direct measurement'' denotes the measurement with a single setting of the measurement apparatus.
%  directly measures a million-dimensional photonic spatial state with a single setting of the measurement apparatus.
Figures~\ref{fig:4}(a) and (b) show the diagram of its time evolution and the optical system used in the experiment.
Here, the transverse spatial mode of a photon and its polarization mode are used as the measured system and the probe, respectively.
The total initial state is set as $\ket{\psi}\ket{\HH}$, where $\ket{\psi}$ gives the wavefunction in the transverse spatial mode as $\psi(x)=\bracketi{x}{\psi}$, and $\ket{\HH}$ is a horizontally polarized state. 
Next, the transverse position of the photon is weakly measured by coupling it to its polarization; this weak interaction is described as $\exp(-\ii\xi\ket{x}\bra{x}\otimes\hat{\sigma}_y)=\hat{1}-\ii\xi\ket{x}\bra{x}\otimes\hat{\sigma}_y+O(\xi^2)$ and realized by a slightly tilted half-wave plate (HWP) at the position $x$.
In the diagram of Fig.~\ref{fig:4}(a), this weak interaction is regarded as the operation that extracts a small portion of $\ket{\psi}$ proportional to $\xi$ from the probe mode $\ket{\HH}$, transfers this portion to the mode $\ket{\VV}$ (vertically polarized state), and multiplies it by $\ket{x}\bra{x}$.
The photon undergoes an optical Fourier transformation induced by the 2-$f$ optical system and then is post-selected in $\ket{p_0}:=\ket{p=0}$ by the pinhole placed in the Fourier plane.
Finally, the photon passes through the polarizer to measure the expectation values of $\hat{\sigma}_x=\ket{\HH}\bra{\VV}+\ket{\VV}\bra{\HH}$ and $\hat{\sigma}_y=-\ii\ket{\HH}\bra{\VV}+\ii\ket{\VV}\bra{\HH}$, from which the weak value $\bracket{\ket{x}\bra{x}}\sub{w}$ is derived experimentally. 
When the pre- and post-selected states are $\ket{\psi}$ and $\ket{p_0}$, the weak value $\bracket{\ket{x}\bra{x}}\sub{w}$ is proportional to the wavefunction $\psi(x)$:
\begin{eqnarray}
\bracket{\ket{x}\bra{x}}\sub{w}
=\frac{\bracketi{p_0}{x}\bracketi{x}{\psi}}{\bracketi{p_0}{\psi}}
\propto\psi(x).
\end{eqnarray}
As described above, weak measurements can be used to directly measure the complex wavefunction $\psi(x)$.
However, when using this method, it is necessary to perform a large number of measurements to achieve high precision because the system--probe interaction is very weak, and the majority of the incoming photons are discarded through post-selection. 
It is also necessary to scan the location $x$ of the HWP for the weak interaction, which requires the measurement time scaling with the dimension of the system's Hilbert space and makes it difficult to characterize the state of a system with large dimensions. 

Next, we show that the direct measurement system of wavefunctions using weak measurement can be modified to develop a scan-free and more efficient measurement method according to the transformation rule of the diagram.
The weak measurement system depicted in Fig.~\ref{fig:4}(a) can be transformed into the system described by the diagram of Fig.~\ref{fig:4}(c).
Here, we select the value of $\xi$ as the maximum value ($\xi=1$) to increase the measurement efficiency.
In order to realize the scan-free direct measurement of wavefunctions, the operators in Fig.~\ref{fig:4}(c) are moved counterclockwise along the loop according to the transformation rule of the diagram; as a result, the diagram depicted in Fig.~\ref{fig:4}(d) is obtained.
Since the measurement basis of the system becomes $\{\ket{x}\bra{x}\}$, wavefunctions can be obtained without scanning by using an appropriate detector with spatial resolution.

The optical system corresponding to the diagram of Fig.~\ref{fig:4}(d) is shown in Fig.~\ref{fig:4}(e).
The total initial state is set as $\ket{\psi}\ket{+}=\ket{\psi}(\ket{\HH}+\ket{\VV})/\sqrt{2}$.
The probe-controlled system transformation $\hat{T}=\ket{p_0}\bra{p_0}\otimes\ket{\HH}\bra{\HH}+\hat{1}\otimes\ket{\VV}\bra{\VV}$ is realized by setting a polarizer pinhole, which transmits all the vertically polarized light, and horizontally polarized light at the center of the pinhole, on the Fourier plane.
After passing through the 2-$f$ optical system, photon's polarization is measured in the bases of $\{\ket{\pm}\}$ (diagonal polarization) and $\{\ket{\!\pm\ii}\}$ (circular polarization), after which the photon enters the detector with spatial resolution such as a charge-coupled device (CCD) camera.
Let $P(\pm)$ and $P(\pm\ii)$ be the probabilities that the initial state undergoes the transformation $\hat{T}$, the system is projected onto $\ket{x}$, and the probe is projected onto $\ket{\pm}$ and $\ket{\!\pm\ii}$, respectively. 
Using $P(\pm)$ and $P(\pm\ii)$, the measured wavefunction is obtained as
\begin{eqnarray}
P(+)-P(-)+\ii[P(+\ii)-P(-\ii)]
&=
\bracketi{\psi}{p_0}\bracketi{p_0}{x}\bracketi{x}{\psi}
\propto\psi(x).\label{eq:5}
\end{eqnarray}
As described above, the wavefunction $\psi(x)$ is directly measured in the optical system of Fig.~\ref{fig:4}(e).

The main advantage of this approach is that the number of measurements conducted for the direct measurement of wavefunctions can be reduced because of the scan-free measurement and its high efficiency.
It should be noted that the scan-free direct measurement of wavefunctions using weak measurement has been realized in a previous study \cite{shi2015scan}.
However, this approach requires a large number of measurements because of the inefficiency of weak measurements, and the measured wavefunction is determined as a reciprocal of the weak value.
In contrast, the technique proposed in this work is more efficient because weak interactions are not used, and wavefunctions can be obtained directly without calculating their reciprocal values, as shown in Eq.~(\ref{eq:5}).

In addition, besides the practical advantage, our approach has a conceptual advantage that the mechanism of wavefunction determination is more straightforward.
In the direct measurement of wavefunctions using weak measurement \cite{lundeen2011direct}, the necessity of weak measurements is not obvious, and it is difficult to understand how wavefunctions can be obtained in the optical system.
On the other hand, weak measurements are no longer required in the proposed method, and the mechanism of wavefunction determination is more clear; that is, in front of the detector depicted in Fig.~\ref{fig:4}(e), the interference between the measured light (vertical polarization) and plane-wave self-reference light (horizontal polarization) leads to the direct measurement of wavefunctions.
This mechanism of wavefunction determination---interference between measured light and plane-wave self-reference light---has been previously used in classical wavefront sensing applications \cite{goto2016reference}.
Hence, the direct measurement of wavefunctions by weak measurements can be considered equivalent to a simple classical measurement technique---complex amplitude measurement using plane-wave self-reference light.

\section{Conclusion}
 
In this study, we proposed a measurement framework of weak values that represents a modified version of the weak measurement using a qubit probe and some other weak-value measurement methods.
In addition, we constructed a diagram that allows to effectively identify complex values measured by the described measurement systems.
We also showed that the new weak value measurement method with a desired function can be systematically derived using the transformation rule of the diagram. 

Finally, it should be noted that the weak value measurement methods included in the proposed framework could not be utilized for all weak measurement applications.
For example, they cannot be used for the precise measurements of the magnitudes of weak system--probe interactions \cite{hosten2008observation,dixon2009ultrasensitive,magana2014amplification,hallaji2017weak} because such interactions are replaced with the probe-controlled system transformations in the measurement framework. 
On the other hand, these measurement methods can be used instead of the weak measurement for studying various fundamental problems in quantum mechanics \cite{PhysRevLett.102.020404,yokota2009direct,lund2010measuring,kocsis2011observing,goggin2011violation,rozema2012violation,denkmayr2014observation,kaneda2014experimental,mahler2016experimental} and direct measurement of complex functions such as wavefunctions and pseudo-probability distributions of the initial state of the system \cite{lundeen2011direct,lundeen2012procedure,salvail2013full,kobayashi2014stereographical,malik2014direct,mirhosseini2014compressive,shi2015scan,thekkadath2016direct,piacentini2016measuring,bolduc2016direct}; it is because the purpose of these applications is to experimentally determine the weak values themselves regardless of the utilized measurement method.
It is expected that comprehensive understanding of various weak value measurement methods yielded by this study will help to derive suitable methods for particular applications and provide intuitive understanding of the elusive concept of weak values.

\section*{Acknowledgements}
This research was supported by JSPS KAKENHI Grant Number 16K17524, the Matsuo Foundation, and the Research Foundation for Opto-Science and Technology.

\section*{References}
% %\bibliography{iopart-num}
\bibliographystyle{iopart-num}
% \bibliography{ref}

\providecommand{\newblock}{}

% \clearpage

% \appendix

% \section{List of macros for formatting text, figures and tables}

\end{document}